\documentclass{article}
\usepackage{amsmath,amssymb,amsfonts,amsthm}
\usepackage{multirow}
\usepackage{color}
    \usepackage{slashbox}
   \usepackage{lscape}
\usepackage{pdflscape}
     \usepackage[all]{xy}
\usepackage[all]{xy}
\definecolor{Light}{gray}{.80}
\definecolor{Dark}{gray}{.20}
\newcommand{\cs}[3]{{{#3} \brace {#1 #2}}}

\newcommand{\h}[1]{\mathop{\hat{\lambda}}\limits_{#1}\ \!\!\!}
\newcommand{\hh}[1]{\mathop{{\lambda}}\limits_{#1}\ \!\!\!}

\newcommand{\li}[1]{\mathop{L}\limits_{#1}\ \!\!\!}

\newcommand{\edf}{\ {\mathop{=}\limits^{\rm def}}\ }

\newcommand{\C}[1]{\mathop{\hat{C}}\limits_{#1}\ \!\!\!}
\begin{document}

\begin{center}{ \Large\bf{An AP-Structure with  Finslerian Flavor: Path Equations} }
\end{center}
\begin{center}
{\Large M. I. Wanas $^{1,4}$, M. E. Kahil $^{2,4}$ and Mona M. Kamal
 $^{3,4}$}

 $^1$Astronomy Department,Faculty of Science, Cairo University, Giza, Egypt

~~Email:{miwanas@sci.cu.edu.eg}

~$^2$The American University in Cairo, New Cairo, Egypt

~~Email:{kahil@aucegypt.edu}

~$^3$Mathematics Department, Faculty of Girls, Ain Shams
University,Cairo, Egypt.

~~E-mail:monamkamal@eun.eg

~$^4$Egyptian Relativity Group (ERG) URL:http://www.erg.eg.net
\end{center}

\abstract {\large The Bazanski approach for deriving paths is applied to Finsler geometry. The approach is generalized and applied to a new developed geometry called "Absolute parallelism with a Finslerian Flavor" (FAP). A sets of path equations is derived for the FAP.  This is the horizontal (h) set. A striking feature appears in this set, that is: the coefficient of torsion term, in the set, jumps by a step of one-half from one equation to the other. This is tempting to believe that the h-set admits some quantum features.  Comparisons  with the corresponding sets in other geometries  are  given. Conditions to reduce the set of path equations obtained, to well known path equations in some geometries are summarized in a schematic diagram.\\
{\bf Keywords}: AP-Geometry; Finsler Geometry; Bazanski approach; Quantum features.\\
{\it PACS: 02.40.Hw}}
{\large
\section{Introduction}

It is well known that, in the context of the  philosophy of geometrization of physics, paths (curves)  in an appropriate geometry are used to represent physical trajectories of test particles in physical space-times. The geometric object which plays an important part in such equations, is the linear connection of the space used. Path equations are usually derived by applying   Euler-Lagrange equations to a certain Lagrangian defined in the context of the geometry used.  Another geometric equation,  important for physical applications, is the path deviation equation which is usually derived by the same method, but using a different Lagrangian function. In Riemannian geometry, the path and path deviation equations are called the geodesic and geodesic deviation equations,  respectively.

Bazanski [4] has suggested an approach to derive the geodesic and
its deviation equations, in Riemannian geometry,  from one single
Lagrangian, which is not the only advantage  of his approach. This
approach can be summarized as follows:
 The Lagrangian function suggested by Bazanski [4], in the context of Riemannian geometry,  has  the form
 \begin{equation}
 L= g_{\mu \nu} U^{\mu} \frac{D \Psi^{\nu}}{Ds},
 \end{equation}
where $g_{\mu \nu}$ is the metric tensor, $U^{\mu}$ is the unit vector tangent to the geodesic, $\Psi^{\nu}$ is the deviation vector,  $s$ is the parameter characterizing the  geodesic and $ \frac{D \Psi^{\nu}}{Ds}\edf \Psi^{\nu}_{~;\alpha}U^\alpha$ with (;) denotes covariant differentiation using Christoffel symbol.
From the above Lagrangian (1), taking the variation   w.r.t $\Psi ^{\alpha}$  Bazanski has obtained,
 \begin{equation}
\frac{dU^{\alpha}}{ds}+ \cs{\mu}{\nu}{\alpha} U^{\mu} U^{\nu}=0 . \end{equation}
% Also, taking the variation w.r.t $U^{\alpha}$ one obtains
%\begin{equation} \frac{D^2 \Psi^{\alpha}}{Ds^{2}} = R^{\alpha}_{\beta \mu \nu} U^{\beta} U^{\mu}\Psi^{\nu},
%\end{equation}
 Equation (2) is the  geodesic  equation of Riemannian geometry.  Varying (1) w.r.to $U^\alpha$ he has obtained the geodesic deviation equation[4], viz
 \begin{equation} \frac{D^2 \Psi^{\alpha}}{Ds^{2}} = R^{\alpha}_{.\beta \mu \nu} U^{\beta} U^{\mu}\Psi^{\nu},
\end{equation}
where $R^{\alpha}_{.\beta \mu \nu}$ is the Riemman-Christoffel curvature tensor and
$$\frac{D^2 \Psi_{\alpha}}{Ds^{2}}\edf \frac{d}{ds}(\frac{D\psi_\alpha}{Ds})-\cs{\mu}{\alpha}{\epsilon}\frac{D\psi_\epsilon}{Ds}U^\mu~.
$$
The importance of the Bazanski approach appears  when it is extended to non Riemannan geometries, that contain  simultaneously non vanishing curvature and torsion [9], [5].  This approach has also been generalized in Riemannian geometry to express the Papapetrou equation for a rotating object and the Dixon equation for a  charged rotating  object together with their corresponding deviation equations [6].\\

  The main advantage of this approach is that it allows one directly to  insert the linear connection, of  a chosen geometry, into the Lagrangian function and consequently into the resulting path and path deviation equations. A new interaction and some quantum features have been discovered by applying this approach to different versions of  absolute parallelism (AP-) geometry [9],[10] and other geometries [5].

The aim of the present work is to apply the Bazanski approach to a version of AP-geometry in which the building blocks of the geometry are functions of both position and direction, the FAP-geometry[2],[7]. The paper is organized in the following manner. A brief review on Finsler geometry and FAP geometry are given in section 2. In section 3, we apply the Bazanski approach in case of Finsler  geometry, for the sake of comparison. In section 4, we  apply the same approach and the conventional one, for  comparison, to the case of FAP-geometry. Comparisons of the results obtained in sections 4 and 3 with those previously obtained in cases of other geometries  are given in section 5 together with some concluding remarks.

Throughout this paper we are going to use the following
notations:\\\\ \\ \\

\begin{center}
{\bf Table 1: List of notations used}
\end{center}

\begin{tabular}{cl}
  \hline
 Symbol & Definition\\\hline
  $x(t)$ & For position.   \\\\
  $y(\edf\frac{dx}{dt})$ & For the direction. \\\\
  $,$ & Ordinary infix partial differential operator with respect to x.  \\\\
   $:$ & Ordinary infix partial differential operator with respect to y.  \\\\
  $\hat{\cs{\beta}{\gamma}{\alpha}}$ &  Cartan linear connection. \\\\
  $\hat{;}$ & Infix covariant differential operator using Cartan connection. \\\\
  $G^{\alpha}_{. \beta}$& Non-linear connection of Finsler geometry.\\\\
  $\delta_\mu$& Prefix partial differential operator using the non-linear connection  $G^{\alpha}_{. \beta}$\\\\
  $N^{\alpha}_{. \beta}$&Non linear connection of the FAP-geometry.\\\\
  $\hat{,}$& Infix partial differential operator using the non linear connection  $N^{\alpha}_{. \beta}$\\\\
   $\hat{\Gamma}^{\alpha}_{.\beta\gamma}$ & Linear connection of the Cartan type of the FAP-geometry. \\\\
  $_{||}$&Infix covariant differential operator using the symmetric part  \\&of the connection $\hat{\Gamma}^{\alpha}_{.\beta\gamma}$.\\\\
 $_{+||}$&Infix covariant differential operator using the connection $\hat{\Gamma}^{\alpha}_{.\beta\gamma}$. \\\\
 $_{-||}$&Infix covariant differential operator using the dual connection \\&  $\widetilde{\hat{\Gamma}}~^{\alpha}_{.\beta\gamma}(\edf\hat{\Gamma}^{\alpha}_{.\gamma\beta})$. \\\\
$|$&Infix covariant differential operator using the tensor $\hat{C}^{\alpha}_{.\beta\gamma}$. \\\\
$\Psi^{\alpha}$ & Deviation vector. \\\\
  $U^{\alpha}$, $y^{\alpha}$ , $\xi^{\alpha} $, $ \eta^{\alpha}$, $\zeta^{\alpha}$ &  Tangent vectors for different paths. \\\\
  $L_{F}$ & Lagrangian of Finsler geometry using the conventional approach. \\\\
  $L_{F(B)}$ & Lagrangian of Finsler geometry using the Bazanski approach. \\\\
  $L_{FAP}$ & Lagrangian of FAP-geometry using the conventional approach. \\\\
  $L_{FAP(B)}$ & Lagrangian of FAP-geometry using the Bazanski approach. \\\\
  \hline
\end{tabular}

\section{Basic Preliminaries }
\subsection{Finsler Space}
Finsler space $(M,F)$ is an n-dimensional differentiable manifold $M$ equipped with a scalar, $F(x,y)$ , function of $x(t)$ and $y (\edf \frac{dx}{dt}= \dot x)$ where $t$ stands for an invariant parameter. $(x,y)$ are defined to be the coordinates on the tangent bundle $TM$. The scalar function  $F(x,y)$ satisfies the following properties \footnote{ For more details on Finsler geometry, the reader is referred to [11] [12]}\\
(a) $F (x,y)$ is  $C^{\infty}$ on $\tau M( = TM \backslash{\{0\}})$. \\
(b)$F (x,y)$ is positively  homogeneous of degree 1 in y. \\
(c) $y ~\epsilon ~ \tau M$ and transforms as $\bar{y}^{\alpha}= \frac{\partial \bar{x}^{\alpha}}{\partial x^{\beta}}y^{\beta}$. \\
(d) The  tensor $\hat{g}_{\alpha \beta}(x,y)$ is defined as %\footnote{Throughout the present work we use the comma $(,)$ for partial differentiation w.r.t. $(x)$%and the colon$(:)$ for differentiation with respect to y.}
,
$$\hat g_{\alpha \beta}(x,y) \edf \frac{\partial^{2}E}{\partial y^{\alpha} \partial y^{\beta}} = E_{: \alpha \beta},$$
 where the colon $(:)$ is to characterize partial differentiation with respect to $y$, and $E$ is the energy of the space defined by
\begin{equation}
 E \edf \frac{1}{2} F^{2},
\end{equation}
and
\begin{equation}
F^{2} \edf \hat g_{\alpha \beta}y^{\alpha}y^{\beta}.
\end{equation}
The tensor $\hat g_{\alpha \beta}$ is the metric of Finsler structure, such that
\begin{equation}
\hat g_{\alpha \gamma}\hat g^{\beta \gamma} = \delta^{\beta}_{~\alpha},
\end{equation}
where $\hat g^{\beta \gamma}$ is the normalized cofactor of $\hat g_{\alpha \beta}$ in the matrix $(\hat g_{\alpha \beta})$. This matrix  is assumed to be non-degenerate and its determinant is denoted by $\hat g$.  These tensors are used for raising and lowering tensor indices.\\

Differentiating $\hat g_{\alpha \beta}$ with respect to $y^{\rho}$ will produce the third order tensor,
\begin{equation}
 \textbf{C}_{\mu \nu \rho} \edf \hat g_{\mu \nu : \rho}= E_{: \mu \nu \rho},
\end{equation}
such that $\textbf{C}_{\alpha \beta \gamma}$ satisfying the following properties: \\
(a) It is symmetric with respect to all indices. \\
(b) It is tensor of type $(0,3)$. \\
(c)It is  positively homogeneous of degree $-1$. \\
Consequently, applying Euler's theorem one can find that,
\begin{equation}
 \textbf{C}_{\alpha \beta \gamma} y^{\alpha} = \textbf{C}_{\alpha \beta \gamma}y^{\beta}=  \textbf{C}_{\alpha \beta \gamma}y^{\gamma}=0.
\end{equation}
{\bf{Theorem 1}}: {\it{A necessary and sufficient condition for a
Finsler space to become a Riemannian one is the identical vanishing
of $C_{\alpha \beta \gamma}$ tensor}}.\\\\ \\ \\ \\
\underline{Non-Linear Connection} \\
Using the metric tensor $\hat g_{\alpha \beta}$ ( function of $(x,y)$), one can  define the geometric object $G^{\alpha}_{. \beta \gamma}$:
\begin{equation} G^{\alpha}_{. \beta \gamma} \edf \frac{1}{2}\hat g^{\alpha \sigma}(\hat g_{\gamma \sigma, \beta }+ \hat g_{\beta  \sigma, \gamma}-\hat g_{\beta \gamma , \sigma}),
\end{equation}
 which is neither a tensor nor a connection.
 This  quantity is used to define the spray,
\begin{equation}
G^{\alpha} \edf G^{\alpha}_{. \beta \gamma} y^{\beta} y^{\gamma}.
\end{equation}
Differentiating $G^{\alpha}$ with respect to $y^{\beta}$ one obtains the non-linear connection
\begin{equation}
G^{\alpha}_{. \beta} \edf  G^{\alpha}_{~:\beta}.
\end{equation}
Using (11), we can define the following differential operator:
\begin{equation}
\delta_{\beta} \edf \partial_{\beta} - G^{\alpha}_{. \beta} \frac{\partial}{\partial y^{\alpha}}.
\end{equation}
\underline{Cartan Linear Connection} \\
The following object
\begin{equation}
 \hat{\cs{\beta}{\gamma}{\alpha}}\edf \frac{1}{2}\hat g^{\alpha \sigma}(\delta_{\beta}\hat g_{\gamma \sigma}+ \delta_{\gamma}\hat g_{\beta  \sigma}-\delta_{\sigma}\hat g_{\beta \gamma})
\end{equation}
is found to be transformed as a linear connection. It is a metric connection called Cartan linear connection. \\\\
\underline{Berwald Linear Connection} \\
Another type of linear connections is defined by
\begin{equation}
 ^{*}{G^{\alpha}_{. \beta \gamma}} \edf \frac {\partial G^{\alpha}_{. \beta}}{\partial y^{\gamma}}=G^{\alpha}_{~~\beta : \gamma}= G^{\alpha}_{~~: \beta \gamma}. \end{equation}
This connection is non-metric having no torsion due to its symmetry with respect to its lower index. This connection is known as Berwald linear connection. \\\\
\underline{The horizontal (h-) derivative} of an arbitrary vector $A_\mu$  is defined by
\begin{equation}
A_ {\alpha} \hat{;}_{ \beta}\edf  \delta_{\beta}A_{\alpha} -  \hat{\cs{\alpha}{\beta}{\mu}}A_{\mu}.
\end{equation}
\underline{The vertical (v-) derivative} of the vector $A_\mu$, is defined by
\begin{equation} A_{{\mu}|_{\nu}} \edf A_{\mu : \nu} - A_{\alpha}\textbf{C}^{\alpha}_{.\mu\nu},
\end{equation}
where $\textbf{C}^{\alpha}_{.\mu \nu}$ is defined as
$$ \textbf{C}^{\alpha}_{.\mu \nu} \edf g^{\alpha \beta}\textbf{C}_{\beta \mu \nu}. \eqno(i) $$
{\bf Path equation in Finsler geometry: The conventional approach}\\
Consider the following Lagrangian [8]
\begin{equation}
L=g_{\mu \nu}y^{\mu} y^{\nu}
\end{equation}
where
$$ y^{\alpha} \edf \frac{dx^\alpha}{dt}, \eqno(17a) $$
and t is parameter characterizing the path.
The variation of (17) with respect to $y^{\alpha}$ leads to the path equation
\begin{equation}
\frac{\bar{D}y^{\alpha}}{\bar{D}t}=0,
\end{equation}
where,
$$
\frac{\bar{D}y^{\alpha}}{\bar{D}t}\edf \frac{d y^{\alpha}}{ d t} + {\hat{\cs{\beta}{\gamma}{\alpha}}} y^{\beta}y^{\gamma},
$$

Equation(18) is the path equation of Finsler space (cf. [12]).
\subsection{A Brief Review of the FAP-geometry}
In the present Subsection, we give a brief summary of the previous work on FAP
already published in [2],[7].

An absolute parallelism space with  Finslerian Flavor  $(M, \li{i} ~)$ is an n-dimensional differentiable manifold $M$ equipped with n-Lagrangians $ \li{i} \equiv \li{i}(x,y) $ functions of the position $(x)$ and the direction $(y)$\footnote{Through out this paper, we are going to use Greek indices for coordinates components and a Latin indices to enumerate functions, and vectors.}, $(i=1,2,3,4....,n)$. The fundamental set of FAP-structure $\li{i}$ satisfies  the following properties [2]: \\
(a) $\li{i} (x,y)$ is  $C^{\infty}$ on $\tau M (= TM \backslash{\{0\}})$ \\
(b) $\li{i} (x,y) > 0$, $y~ \epsilon ~\tau M ,  y = \dot x (= \frac{dx}{dt})$ \\
(c)$\li{i} (x,y)$ is positively  homogeneous of degree 1.\\
(d) The vector fields
\begin{equation}
\h{i}_{\mu} \edf \frac{\partial \li{i}}{\partial y^{\mu}}= {\li{i}_{: \mu}},
\end{equation}
are the building blocks of FAP-geometry.
The objects defined by (19) are assumed to be linearly independent . It can be shown that ${{\h{i}}}{_{\mu}}$, considering the index $\mu$, is transformed as components of covariant vectors under diffeomorphism. Consequently, using Euler's theorem we can write,
\begin{equation}
\li{i} \edf {\h{i}}{_{\mu}} y^{\mu}.
\end{equation}
{\bf{Theorem 2}:} {\it A necessary and sufficient condition for an FAP-structure to be an AP-structure is the vanishing of the type $(0,2)$ tensor given by [2]}
\begin{equation}
\C{i}_{\mu \nu}  \edf \frac{\partial \h{i}_{\mu}}{\partial y^{\nu}} = \li{i}_{: \mu \nu},
\end{equation}
which is  symmetric w.r.t. $(\mu, \nu) $ . Consequently, by using Euler's theorem we can write,
\begin{equation}
 \C{i}_{\mu \nu}y^{\mu}=\C{i}_{\mu \nu}y^{\nu} =0 .
\end{equation}
Using (21) we can define the tensor \begin{equation}
\hat{C}^{\alpha}_{.\mu\nu}\edf \h{i}^{\alpha} \C{i}_{\mu\nu}.\end{equation}

The following relations can be easily verified,
\begin{equation}
 \h{i}^{\mu} \h{i}_{\nu} = \delta^{\mu}_{\nu},
\end{equation}
\begin{equation}
\h{i}^{\mu} \h{j}_{\mu} = \delta_{ij},\end{equation}
where $\h{i}^{\mu}$ are the normalized cofactor of $\h{i}_{\mu}$ in the matrix ($\h{i}_{\mu}$).
Using the building blocks of the FAP-structure one can define the non-linear connection,
\begin{equation}
N^{\nu}_{. \mu} \edf y^{\alpha} \h{i}^{\nu} \h{i}_{\alpha,\mu} . \end{equation}
Also, one can define the following linear  connection of the Cartan type,
\begin{equation}\label{ct}
 \hat\Gamma^{\alpha}_{. \mu \nu} \edf \h{i}^{\alpha}\h{i}_
 {\mu \hat{,} \nu} ,
 \end{equation}
where the comma with a hat is an infix operator defined for any arbitrary vector field   $A_{\mu} $ as
\begin{equation}
 A_{\mu \hat{,} \nu} \edf A_{\mu, \nu} - A_{\mu : \beta} N^{\beta}_{. \nu} . \end{equation} It is to be considered that all objects given, so far, are defined in terms of the building blocks of the FAP-structure. \\
 Another  linear connection of the Bervald type is defined by,
  \begin{equation}\label{10}
  \hat{B}^{\mu}_{.~ \alpha \beta} \edf N^{\mu}_{~\alpha : \beta} =
\frac{\partial}{\partial y^{\beta}} (y^{\nu} \h{i}^{\mu}\h{i}_{\nu,~
\alpha}).\end{equation}

For the FAP-structure we have a set of, at least, four d-connections: $D=(\hat{\cs{\mu}{\nu}{\alpha}}, N^\alpha_{~\mu}$, $\hat{C}^\alpha_{.\mu\nu})$, $ \hat{D}=(\hat{\Gamma}^\alpha_{.\mu\nu},N^\alpha_{~\mu}, \hat{C}^\alpha_{.\mu\nu})$, $
\widetilde{D}=(\widetilde{\hat{\Gamma}}~^\alpha_{.\mu\nu}, N^\alpha_{~\mu}, \hat{C}^\alpha_{.\mu\nu}) $ and
$\overline{D}=(\overline{\hat{\Gamma}}~^\alpha_{.\mu\nu}, N^\alpha_{~\mu}, \hat{C}^\alpha_{.\mu\nu}).$ \\

{\underline{h-derivatives}}\\
Using the Cartan-type connection (27), its dual and its symmetric part,
 one can define the following horizontal (h-) derivatives:
\begin{equation}
A^{\stackrel{\mu}{+}} _{.~{||} \nu} (x,y) \edf A^{\mu} _{~\hat{,} \nu} (x,y) +
A^{\alpha}{\hat\Gamma^{\mu}_{.\alpha \nu}} (x,y), \end{equation}
\begin{equation}
A^{\stackrel{\mu}{-}} _{.~{||} \nu}(x,y) \edf A^{\mu} _{~\hat{,}
\nu} (x,y)+ A^{\alpha}\widetilde{\hat{\Gamma}}~^\mu_{.\alpha \nu}
(x,y),
\end{equation}
\begin{equation}
A^{\mu}_{\stackrel{.||\nu}{~~o}} (x,y) \edf A^{\mu} _{~\hat{,} \nu} (x,y)+
A^{\alpha}{\hat\Gamma^{\mu}_{.( \alpha \nu )} } (x,y),
\end{equation}
where the double stroke and $(+)$ sign denote tensor derivative using linear
connection (27) and the double stroke and the $(-)$ sign denote the
tensor derivative using the dual connection,
\begin{equation}
{{\widetilde{\hat{\Gamma}}~^{\alpha}_{. \mu \nu}} }(x,y) \edf
{\hat\Gamma^{\alpha}_{. \nu \mu}}(x,y),
\end{equation}
 while  the double stroke alone
characterizes tensor derivatives using the symmetric connection,
\begin{equation}
{\hat\Gamma^{\alpha}_{. ( \mu \nu )}} (x,y) \edf
\frac{1}{2}({\hat\Gamma^{\alpha}_{. \mu \nu}} (x,y) + {\hat\Gamma^{\alpha}_{. \nu
\mu}} (x,y)).
\end{equation}
This may be written as
\begin{equation}
{\hat\Gamma^{\alpha}_{. ( \mu \nu )}}(x,y) \edf \hat{\cs{\mu}{\nu}{\alpha}
}(x,y)+
\frac{1}{2} {\hat\Delta^{\alpha}_{. \mu \nu}}(x,y),
\end{equation}
where,
\begin{equation}
{\hat\Delta^{\alpha}_{. \mu \nu}(x,y)} \edf {\hat\gamma^{\alpha}_{. \mu \nu}}(x,y) +
{\hat\gamma^{\alpha}_{. \nu \mu}}(x,y) ,
\end{equation}
\begin{equation}
{\hat\gamma^{\alpha}_{. \mu \nu}}(x,y) \edf \h{i}^{\alpha}\h{i}_
 {\mu \hat{;} \nu}.
\end{equation}
The operator ($\hat ;$) is used to characterized covariant differentiation in the FAP-geometry, as used in (15).
As a consequence of using the connection (27), it is easy to show
that,
\begin{equation}\label{14}
\h{i}_{\stackrel{\mu}{+} {||} \nu}(x,y) = 0,
\end{equation}
which is the horizontal AP-condition.\\

{\underline{v-derivatives}}\\
Due to the symmetry of $\hat{C}^\alpha _{ .\beta \gamma}$ we have only one v-derivative viz
\begin{equation}
A^{\mu}|_{ \nu} (x,y)\edf A^{\mu} _{~: \nu} + A^{\alpha}\hat{C}^{\mu}_{.\alpha \nu},
\end{equation}
where long stroke  denotes tensor derivative using (23).
 Accordingly, we get
\begin{equation}
\h{i}_{\mu} |_{\nu} (x,y)= 0,
\end{equation}
which is the vertical AP-condition. \\
%{\bf{Theorem 3}}: {\it{A necessary and sufficient condition for FAP-space to become an AP-space is the identical vanishing of its $\C{i}_{\alpha \beta}$ tensor}}[2]. \\
{\bf{Theorem 3}}: {\it{In case of $\hat{C}_{[\alpha\beta] \gamma}=0$, the FAP-space has an associated Finsler Space having  the metric[2],}}
\begin{equation}\label{15}
\hat g_{\mu \nu} \edf \h{i}_{\mu}\h{i}{_{\nu}},
\end{equation}
\begin{equation}\label{16}
\hat g^{\mu \nu} \edf \h{i}^{\mu}\h{i}{^{\nu}}.
\end{equation}

It is clear from the definition (41) and (38) that

\begin{equation}\label{17}
\hat{g}_{\stackrel {\mu \nu|| \sigma}{++~~~} } \equiv0,\end{equation}
which means that the linear connection (\ref{ct}) is a metric one. Also, it can be shown, using (40) and (42), that
\begin{equation}\label{18}
\hat{g}_{\mu \nu}|_{\sigma} \equiv 0.\end{equation}
Relations (43) and (44) mean simply that the operation of raising and lowering tensor indices commutes with the operations of the positive h- and v- differentiations.

The torsion of the linear connection (27) is defined by,
\begin{equation}\label{38}
\hat{\Lambda}^\alpha_{.
\mu\nu}\edf\hat{\Gamma}^{\alpha}_{.~ \mu \nu}-\hat{\Gamma}^{\alpha}_{.~\nu \mu }=\hat{\gamma}^{\alpha}_{.~ \mu \nu}-\hat{\gamma}^{\alpha}_{.~\nu \mu },\end{equation}
which gives, by contraction, the vector
\begin{equation}\label{39} \hat{c}_\mu\edf\hat{\Lambda}^\mu_{.\alpha\mu}=\hat{\gamma}^\mu_{.\alpha\mu}.\end{equation}
Also, the torsion of the {\it the Bervard-type} connection (29) is defined by,
\begin{equation}\label{42} \hat{T}^\alpha_{.
\mu\nu}\edf\hat{B}^{\alpha}_{.~ \mu \nu}-\hat{B}^{\alpha}_{.~\nu \mu },\end{equation}
 which gives the vector,
\begin{equation}\label{43} \hat{T}_{
\mu}\edf\hat{T}^\alpha_{.
\mu\alpha}.\end{equation}

 Commutation relations between the above mentioned derivatives give the following curvature tensors corresponding to the connections (27), (33), (34) , (13) and the object (23), respectively,

\begin{equation}\label{19}
\hat{B}^\mu_{.\alpha\beta\sigma}\edf\hat{\Gamma}^{\mu}_{. \alpha \sigma\hat{,}\beta}-\hat{\Gamma}^{\mu}_{. \alpha\beta\hat{,}\sigma}+\hat{\Gamma}^{\mu}_{. \epsilon \beta}\hat{\Gamma}^{\epsilon}_{. \alpha\sigma}-\hat{\Gamma}^{\mu}_{. \epsilon \sigma}\hat{\Gamma}^{\epsilon}_{. \alpha\beta}-\hat{C}^\mu_{.\alpha\epsilon}N^\epsilon_{.\beta\sigma},~~~~~~~~~~~~~~~~
\end{equation}

\begin{equation}\label{34}
\widetilde{\hat{B}}~^\mu_{.\alpha\beta\sigma}\edf\widetilde{\hat{\Gamma}}~^{\mu}_{. \alpha \sigma\hat{,}\beta}-\widetilde{\hat{\Gamma}}~^{\mu}_{. \alpha\beta\hat{,}\sigma}+\widetilde{\hat{\Gamma}}~^{\mu}_{. \epsilon \beta}\widetilde{\hat{\Gamma}}~^{\epsilon}_{. \alpha\sigma}-\widetilde{\hat{\Gamma}}~^{\mu}_{. \epsilon \sigma}\widetilde{\hat{\Gamma}}~^{\epsilon}_{. \alpha\beta}-\hat{C}^\mu_{.\alpha\epsilon}N^\epsilon_{.\beta\sigma},~~~~~~~~~
\end{equation}

\begin{equation}\label{35}
\overline{\hat{B}}~^\mu_{.\alpha\beta\sigma}\edf\hat{\Gamma}^{\mu}_{. (\alpha \sigma)\hat{,}\beta}-\hat{\Gamma}^{\mu}_{. (\alpha\beta)\hat{,}\sigma}+\hat{\Gamma}^{\mu}_{. (\epsilon \beta)}\hat{\Gamma}^{\epsilon}_{. (\alpha\sigma)}-\hat{\Gamma}^{\mu}_{. (\epsilon \sigma)}\hat{\Gamma}^{\epsilon}_{. (\alpha\beta)}-\hat{C}^\mu_{.\alpha\epsilon}N^\epsilon_{.\beta\sigma},~~~~~
\end{equation}

\begin{equation}\label{37}
\hat{R}^\mu_{.\alpha\beta\sigma}\edf\hat{\cs{\alpha}{\sigma}{\mu}}_{\hat{,}\beta}-\hat{\cs{\alpha}{\beta}{\mu}}_{\hat{,}\sigma}+\hat{\cs{\epsilon}{\beta}{\mu}}\hat{\cs{\alpha}{\sigma}{\epsilon}} -\hat{\cs{\epsilon}{\sigma}{\mu}}\hat{\cs{\alpha}{\beta}{\epsilon}}
-\hat{C}^\mu_{.\alpha\epsilon}N^\epsilon_{.\beta\sigma},~
\end{equation}
and
\begin{equation}\label{21}
\hat{S}^\mu_{.\alpha\beta\sigma}\edf \hat{C}^{\mu}_{. \alpha \sigma:\beta}-\hat{C}^{\mu}_{. \alpha \beta:\sigma}+\hat{C}^{\mu}_{. \epsilon \beta}\hat{C}^{\epsilon}_{. \alpha \sigma}-\hat{C}^{\mu}_{. \epsilon \sigma}\hat{C}^{\epsilon}_{. \alpha\beta}.~~~~~~~~~~~~~~~~~~~~~~~~~~~~~~
\end{equation}\\
Note that Cartan linear connection (13) can be defined in the context of the FAP-geometry using theorem 3.\\

{\bf Theorem 4.} {\it "The horizontal curvature (\ref{19}) vanishes identically"} i.e. $$ \hat{B}^\mu_{.\alpha\beta\sigma}\equiv0 .$$

This theorem can be proved using definitions (27) and (38). \\

{\bf Theorem 5.} {\it "The vertical curvature (53) "vanishes identically"} i.e. $$\hat{S}^\mu_{.\alpha\beta\sigma}\equiv0 .$$

This theorem can be easily proved using definition (23) and (40).\\

The curvature tensor $\hat{B}^\mu_{.\alpha\beta\sigma}$  as defined by (49) can be expressed in terms of the  contortion tensor (37) and curvature (52) i.e.
\begin{equation}\label{49}
\hat{B}^\mu_{.\alpha\beta\sigma}=\hat{R}^\mu_{.\alpha\beta\sigma}+\hat{Q}^\mu_{.\alpha\beta\sigma}
\end{equation}where,
\begin{equation}\label{50}\hat{Q}^\mu_{.\alpha\beta\sigma}\edf\hat{\gamma}^\mu_{\stackrel{.\alpha\sigma||\beta}{~~~~+}}-\hat{\gamma}{ \stackrel{\mu}{+}}_{\stackrel{.\alpha\beta||\sigma}{+-~}}-\hat{\gamma}^\nu_{.\alpha\sigma}\hat{\gamma}^\mu_{.\nu\beta}
  +\hat{\gamma}^\nu_{.\alpha\beta}\hat{\gamma}^\mu_{. \nu \sigma}. \end{equation}
   Now from the definitions (55) and (52), it is clear that neither $\hat{Q}^\mu_{.\alpha\beta\sigma}$ nor $ \hat{R}^\mu_{.\alpha\beta\sigma}$ vanish, while their sum, given by (54), vanishes (Theorem 4). It is to be considered that, the curvature tensor $\hat{R}^\mu_{.\alpha\beta\sigma}$ is  made of the linear connection $\hat{\cs{\alpha}{\beta}{\mu}} $ and the tensors $\hat{C}^{\mu}_{. \alpha\beta}$, $N^{\mu}_{. \alpha\beta}$ while the tensor $\hat{Q}^\mu_{.\alpha\beta\sigma}$ is purely made of the torsion (or the contortion) of the Catan-type connection (27). The addition of these two tensors gives rise to the vanishing of the total curvature of the space (Theorem 4). For the above mentioned comments, we call $\hat{Q}^\mu_{.\alpha\beta\sigma}$ the \underline{\textit{"additive inverse of the curvature tensor"}}  or the \textit{\underline{"anti-curvature"}} tensor.\\

 Another important tensor, of type [1,3], can be defined in the context of FAP-geometry, called the W-tensor. Different versions of this tensor have been defined using the above mentioned differential operators, viz

\begin{equation}\label{054} {\hat W}~^\sigma_{.\mu\alpha\beta}\edf\h{i}^\sigma ( \h{i}_{\stackrel{\mu||\alpha\beta}{+~~~}}-\h{i}_{\stackrel{\mu||\beta\alpha}{+~~~}})  \end{equation}

%\begin{equation}\label{055}  \widetilde{\hat W}~^\epsilon_{.\mu\alpha\beta}\edf \hat{\Lambda}^\epsilon_{\stackrel{.\mu\alpha||\beta}{~~~~-}}-\hat{\Lambda}^\epsilon_{\stackrel{.\mu\beta||\alpha}{~~~~-}}+\hat{\Lambda}^s_{.\mu\beta}\hat{\Lambda}^\epsilon_{.\alpha s}-\hat{\Lambda}^s_{.\mu\alpha}\hat{\Lambda}^\epsilon_{.\beta s}\end{equation}
\begin{equation}\label{055}  \widetilde{\hat W}~^\sigma_{.\mu\alpha\beta}\edf\h{i}^\sigma( \h{i}_{\stackrel{\mu||\alpha\beta}{-~~~}}-\h{i}_{\stackrel{\mu||\beta\alpha}{-~~~}})\end{equation}

%\begin{equation}\label{47}  \overline{\hat W}~^\epsilon_{.\mu\alpha\beta}\edf \frac{1}{2}(\hat{\Lambda}^\epsilon_{.\mu\alpha||\beta}-\hat{\Lambda}^\epsilon_{.\mu\beta||\alpha})+\frac{1}{4}(\hat{\Lambda}^s_{.\mu\beta}\hat{\Lambda}^\epsilon_{.\alpha s}-\hat{\Lambda}^s_{.\mu\alpha}\hat{\Lambda}^\epsilon_{.\beta s})\end{equation}
\begin{equation}\label{47}  \overline{\hat W}~^\sigma_{.\mu\alpha\beta}\edf\h{i}^\sigma ( \h{i}_{{\mu||\alpha\beta}}-\h{i}_{{\mu||\beta\alpha}})\end{equation}

%\begin{equation}\label{48}  \stackrel o {\hat W}~^\epsilon_{.\mu\alpha\beta}\edf \hat{\gamma}^\epsilon_{\stackrel{.\mu\alpha||\beta}{~~~~+}}-\hat{\gamma}^\epsilon_{\stackrel{.\mu\beta||\alpha}{~~~~+}}+\hat{\gamma}^s_{.\mu\beta}\hat{\gamma}^\epsilon_{.s\alpha}
%  -\hat{\gamma}^s_{.\mu\alpha}\hat{\gamma}^\epsilon_{. s \beta}+\gamma^\epsilon_{. \mu s} \hat{\Lambda}^s_{.\alpha \beta}\end{equation}
\begin{equation}\label{48}  \stackrel o {\hat W}~^\sigma_{.\mu\alpha\beta}\edf\h{i}^\sigma ( \h{i}_{{\mu\hat{;}\alpha\beta}}-\h{i}_{{\mu\hat{;}\beta\alpha}})\end{equation}

\begin{equation}\label{} \stackrel \diamond{\hat W}~^\sigma_{.\mu\alpha\beta}\edf\h{i}^\sigma (
\h{i}_{\mu}|_{\alpha\beta}-\h{i}_{\mu}|_{\beta\alpha})\end{equation}

{\bf Theorem 6.}
{\it The curvature and W-tensors have the following coincidences:
  \begin{description}
   \item[(i)]  $ \overline{\hat W}~^\sigma_{.\mu\alpha\beta}$ coincides with the curvature tensor  $\overline{\hat B}~^\sigma_{.\mu\alpha\beta}$.
   \item[(ii)] $\stackrel o {\hat W}~^\sigma_{.\mu\alpha\beta} $ coincides with the curvature tensor $\hat R^\sigma_{.\mu\alpha\beta}$.
 \end{description}}
For more details about the anti-curvature and the W-tensor, the reader is referred to [7].
\section{Path Equations in Finsler-geometry: The Bazanski Approach}
 In this section, for the sake of comparison, we apply  the Bazanski approach to Finsler geometry, in order to obtain its set of path equations using  horizontal (h-)  derivative. \\ Let us  suggest the following appropriate Lagrangian, considering (1), (17),
\begin{equation}
L_{F(B)} \edf g_{\mu \nu} U^{\mu}  \frac{D \Psi^{\nu}}{Ds}
\end{equation}
where $\Psi^{\nu}$  is the deviation vector, $s$ is the parameter of the path and
\begin{equation}
\frac{D \Psi^{\nu}}{Ds} \edf \Psi^{\nu}\hat{;}_{ \alpha}U^{\alpha},
\end{equation}
is the covariant differentiation, with respect to a parameter $s$,
for the deviation vector.
  Using (62) and considering (15) we can expand (61) to take the form:
  $$ L_{F(B)}=  (\delta_{\alpha} \Psi^{\nu} + \Psi^{\beta} \hat{\cs{\alpha}{\beta}{\nu}} ) U^{\alpha} U^{\mu}g_{\mu \nu} ,$$  with $\delta_{\alpha}$ defined by (12). The above expression can be written as
 $$ L_{F(B)}=  g_{\mu \nu} U^{\mu}( \dot{\Psi}^{\nu} +\hat{\cs{\alpha}{\beta}{\nu}} U^{\alpha} \Psi^{\beta} ) , \eqno{(61a)}$$
 where $\dot{\Psi}^{\nu} \edf \frac{d\Psi^{\nu}}{ds}$. Taking the variation of the (61a) with respect $\Psi^{\sigma}$, we get
$$\frac{\partial L_{F(B)}}{\partial \Psi^\sigma}=\hat{g}_{\mu\nu}U^\mu U^\alpha \hat{\cs{\alpha}{\sigma}{\nu}},$$
$$\frac{d}{ds}(\frac{\partial L_{F(B)}}{\partial \dot{\Psi}^\sigma})=\hat{g}_{\mu\sigma\hat{,}\nu}U^\mu U^\nu+\hat{g}_{\mu\sigma}\frac{d U^\mu}{ds}.$$
Substituting from the above two relations into the Euler-Lagrange equation,
$$ \frac{d}{ds}(\frac{\partial L_{F(B)}}{\partial \dot \Psi^{\sigma}}) - \frac{\partial L_{F(B)}}{\partial \Psi^{\sigma}}= 0,
$$
we get after some reductions (note that the Cartan linear connection is metric i.e. $g_{\alpha \beta \hat{;} \gamma} =0$  ) , the path equation
\begin{equation}
{\frac{dU^\mu}{ds}} + \hat{\cs{\alpha}{\beta}{\mu}} U^\alpha U^\beta =
0. \end{equation}
where $\hat{\cs{\mu}{\nu}{\rho}}$ and $\textbf{C}^{\rho}_{. \mu \nu }$ is defined by (13) and ({\it{i}}) respectively , and  $U^{\alpha} \edf \frac{dx^{\alpha}}{ds}$.
  So, equation (63)  is the  path equation  in Finsler geometry obtained using the Bazanski approach, with $U^{\alpha}$ is the tangent of the path and $ (s)$  is its parameter.  This  equation is identical to (18) obtained using the conventional approach.
However, the variation (61a) with respect to $U^{\alpha}$  leads to
the corresponding set of deviation equations, which will be
discussed in our future work.
\section{Path Equations in FAP-geometry}
In the present section, we derive the path equations of the FAP-geometry using two approaches, the conventional approach and the Bazanski one. This is done to get the
corresponding equations and to compare the two approaches.
\subsection{The Conventional Approach}
Modifying (17) to be appropriate for the FAP-geometry, we suggest the following Lagrangian
\begin{equation}
L_{\tiny FAP}=\h{i}_{\mu}\h{i}_{\nu}y^{\mu} y^{\nu},
\end{equation}
where  $y^\alpha$ is given by (17a) and $t$ is the parameter varying along the path. Now, the derivatives of FAP necessary for the variation may be written as

%where\begin{equation}
 %\frac{\delta y^{a}}{\delta t}\edf \end{equation}
$$\frac{\partial L}{\partial x^\sigma}=(\h{i}_{\mu,\sigma}\h{i}_{\nu}+\h{i}_{\mu}\h{i}_{\nu,\sigma})y^\mu y^\nu,$$
$$\frac{\partial L}{\partial y^\sigma}=2\hat{g}_{\mu\sigma}y^\mu,$$
$$\frac{d}{dt}(\frac{\partial L}{\partial y^\sigma})=2\hat{g}_{\mu\sigma,\nu}y^\nu y^\mu+2\hat{g}_{\mu\sigma}\frac{d y^\mu}{dt},$$
note that, $$\frac{d}{dt}=\frac{\partial}{\partial x^\alpha}y^\alpha+\frac{\partial}{\partial y^\alpha}\dot{y}^\alpha.$$
Substituting the above derivatives into the the Euler-Lagrange equation
$$ \frac{d}{dt}(\frac{\partial L_{FAP}}{\partial  y^{\sigma}}) - \frac{\partial L_{FAP}}{\partial x^{\sigma}}= 0,
$$
we get after some reductions (note that the linear connection $\hat\Gamma^{\mu}_{ . \alpha \beta} $ is metric and consequently the identity (43) holds.)
\begin{equation}
{\frac{dy^\mu}{dt}} + \hat\Gamma^{\mu}_{. \alpha \beta}~ y^\alpha y^\beta =
0. \end{equation}

It is to be noted that the appearance of the linear connection $(\hat{\Gamma}^\mu_{. \alpha \beta})$ in the path equation is a result of the variation and the existence of $\h{i}$~  in the Lagrangian (64). This connection does not explicitly written in the Lagrangian. This point will be more discussed later.

\subsection{The Bazanski Approach}
Considering (1) and (64) we may generalize the Bazanski Lagrangian for the FAP-geometry in the following form
 \begin{equation}\label{}
{L_{FAP}}_{(B)}= \h{i}_{\mu}\h{i}_{\nu} M^{\mu}\frac{D
\Psi^{\nu}}{D\tau},
\end{equation}
where $\Psi ^\nu$ is the deviation vector,  $ M^\alpha\ (\edf
\frac{dx^\alpha}{d\tau}) $ is the the unit tangent to the path and
$(\tau)$ is a general parameter varying along the path.

Now, the linear connection (27) is non-symmetric as stated above, so
we have, at least  four linear connections, of the Cartan type, viz
$\hat{\Gamma}^\alpha_{.\beta\gamma}$,
$\hat{\Gamma}^\alpha_{.(\beta\gamma)}$,
$\hat{\widetilde{\Gamma}}~^\alpha_{.\beta\gamma}$  and
$\hat{\cs{\beta}{\gamma}{\alpha}}$. Let us examine the consequences
of using each connection in the definition of $\frac{D
\Psi^{\nu}}{D\tau} $ in (66).
\begin{description}
  \item[(i)] \underline{The use of $\hat{\Gamma}^\alpha_{.\beta\gamma}$}\\
  In this case we write  the Lagrangian (66) in the form,
   \begin{equation}\label{}
{L_{FAP}}_{\stackrel{(B)}{+}}= \h{i}_{\mu}\h{i}_{\nu}
\xi^{\mu}\frac{D \Psi^{\nu}}{D\tau^{+}}
\end{equation}
where $ \xi^\alpha(\edf \frac{dx^\alpha}{d\tau^+} )$ is the tangent
to the path, $(\tau^{+})$ is its parameter and
 \begin{equation}\label{}
 \frac{D \Psi^{\nu}}{D\tau^+} \edf \Psi^{\nu}_{\stackrel{.||\alpha}{~~+}}\xi^{\alpha}.\end{equation}
Using (68) and $$\Psi^{\nu}_{\stackrel{.|| \alpha}{~~+}} \edf
\Psi^{\nu}_{. \hat{,} \alpha} +
\Psi^{\beta}~\hat{\Gamma}^{\nu}_{\beta \alpha} ,  $$ we can expand
the Lagrangian (67) in the form
$$
L_{FAP_{\stackrel{(B)}{+}}} = \h{i}_{\mu}\h{i}_{\nu} \xi^{\mu} (
\dot\Psi^{\nu} + \Psi^{\beta}\xi^{\alpha}~ \hat{\Gamma}^\nu_{. \beta
\alpha})  \eqno{(67a)}
$$ with $$\dot\Psi^{\nu} \edf \Psi^{\nu}_{. \hat{,} \alpha} \xi^{\alpha}.$$ Performing the variation for the Lagrangian (67a) with respect to $\Psi^{\sigma}$ we get the following derivatives
$$\frac{\partial L}{\partial \psi^\sigma}=\hat{g}_{\mu\nu}\xi^\mu \xi^\alpha \hat{\Gamma}^\nu_{.\sigma\alpha},$$
$$\frac{\partial L}{\partial \dot{\psi}^\sigma}=\hat{g}_{\mu\sigma}\xi^\mu,$$
$$\frac{d}{d\tau^+}(\frac{\partial L}{\partial \dot{\psi}^\sigma})=\hat{g}_{\mu\sigma\hat{,}\nu}\xi^\mu \xi^\nu+\hat{g}_{\mu\sigma}\frac{d \xi^\mu}{d\tau^+}.$$

Using the above derivatives in the Euler-Lagrange equation,
$$ \frac{d}{d\tau^{+}}(\frac{\partial L_{FAP(B)}}{\partial  \dot{\Psi}^{\sigma}})- \frac{\partial L_{FAP(B)}}{\partial \Psi^{\sigma}}= 0,
$$
we get after some rearrangements
\begin{equation}
\hat g _{\mu\sigma} \frac{d{\xi}^{\mu}}{d\tau^{+}}+ \hat g _{\nu\sigma}\hat{\Gamma}^\nu_{.\beta\gamma}
\xi^{\beta}\xi^{\gamma} =0,
 \end{equation}
which gives
\begin{equation}
 \frac{d{\xi}^{\alpha}}{d\tau^{+}}+ {\hat{\cs{\mu}{\nu}{\alpha}}}
\xi^{\mu}\xi^{\nu} =- \hat\Lambda_{\mu
 \nu}^{..~\alpha} \xi^{\mu}\xi^{\nu}
 \end{equation}
Equation (70) is the path equation corresponds to the linear
connection $\hat{\Gamma}^{\mu}_{.\alpha \beta}$, and characterized
by the scalar parameter $\tau^{+}$ and the tangent $\xi^{\mu}$.
  \item[(ii)] \underline{The use of $\hat{\Gamma}^\alpha_{.(\beta\gamma)}$}\\
Using the definition of the symmetric part of the linear connection
 (34), we write (66) in the form
   \begin{equation}\label{}
{L_{FAP}}_{\stackrel{(B)}{o}}= \h{i}_{\mu}\h{i}_{\nu} \eta^{\mu}\frac{D \Psi^{\nu}}{D\tau^o}
\end{equation}
 where $ \eta^\alpha(\edf  \frac{dx^\alpha}{d\tau^o} ) $ is the tangent to the path whose parameter is $\tau^o$  and
 \begin{equation}\label{}
 \frac{D \Psi^{\nu}}{D\tau^o} \edf \Psi^{\nu}_{\stackrel{.||\alpha}{~~o}}\eta^{\alpha}\end{equation}
Using (72) and $$\Psi^{\nu}_{\stackrel{.|| \alpha}{~~o}} \edf \Psi^{\nu}_{. \hat{,} \alpha} + \Psi^{\beta}~\hat\Gamma^{\nu}_{(\beta \alpha)} ,  $$ we can expand the Lagrangian (71) in the form
$$
L_{FAP_{\stackrel{(B)}{o}}} = \h{i}_{\mu}\h{i}_{\nu} \eta^{\mu} ( \dot\Psi^{\nu} + \Psi^{\beta}\eta^{\alpha}~ \hat{\Gamma^{\nu}_{. (\alpha \beta)}}) \eqno{(71a)}
$$ with $$\dot\Psi^{\nu} \edf \Psi^{\nu}_{. \hat{,} \alpha} \eta^{\alpha}.$$
Now the variation of (71a) with respect to $\Psi^{\sigma}$ gives the following derivatives,
 $$\frac{\partial {L_{FAP}}_{\stackrel{(B)}{o}}}{\partial \psi^\sigma}=\hat{g}_{\mu\nu}\eta^\mu \eta^\alpha \hat{\Gamma}^\nu_{.(\sigma\alpha)},$$
$$\frac{d}{d\tau^o}(\frac{\partial {L_{FAP}}_{\stackrel{(B)}{o}}}{\partial \dot{\psi}^\sigma})=\hat{g}_{\mu\sigma\hat{,}\nu}\eta^\mu \eta^\nu+\hat{g}_{\mu\sigma}\frac{d \eta^\mu}{d\tau^o}.$$

 Using these derivatives in the Euler-Lagrange equation
 $$ \frac{d}{d\tau^{o}}(\frac{\partial{L_{FAP}}_{\stackrel{(B)}{o}} }{\partial  \dot{\Psi}^{\sigma}})- \frac{\partial{L_{FAP}}_{\stackrel{(B)}{o}} }{\partial \Psi^{\sigma}}= 0,
$$ we get after necessary reductions
 \begin{equation}
 \frac{d{\eta}^{\alpha}}{d\tau^o}+ {\hat{\cs{\mu}{\nu}{\alpha}}}
 \eta^{\mu}\eta^{\nu} =- \frac{1}{2}\hat\Lambda_{\mu
 \nu}^{..~\alpha} \eta^{\mu}\eta^{\nu}.
 \end{equation}
Equation (73) is the path equation corresponds to the linear
symmetric connection $\hat\Gamma^{\alpha}_{(\beta \gamma)}$ and
characterized by the parameter $\tau^{o}$ and the tangent
$\eta^{\alpha}$.
  \item[(iii)]\underline{The use of $\widetilde{\hat{\Gamma}}~^\alpha_{.\beta\gamma}$}\\
 Similarly, in this case, we write the Lagrangian (66) as
   \begin{equation}\label{}
{L_{FAP}}_{\stackrel{(B)}{-}}= \h{i}_{\mu}\h{i}_{\nu} \zeta^{\mu}\frac{D \Psi^{\nu}}{D\tau^-}
\end{equation}
 where $\zeta^\alpha(\edf \frac{dx^\alpha}{d\tau^-}) $ is the tangent to the path whose parameter is $\tau ^-$ and
 \begin{equation}\label{}
 \frac{D \Psi^{\nu}}{D\tau^-} \edf \Psi^{\nu}_{\stackrel{.||\alpha}{~~-}}\zeta^{\alpha}.\end{equation}
Using (75) and
$$\Psi^{\nu}_{\stackrel{.|| \alpha}{~~-}} \edf \Psi^{\nu}_{. \hat{,} \alpha} + \Psi^{\beta}~\widetilde{\hat\Gamma}~^{\nu}_{\beta \alpha} ,  $$
the Lagrangian (74) can be written in the form
 $${L_{FAP}}_{\stackrel{(B)}{-}}= \h{i}_{\mu}\h{i}_{\nu} \zeta^{\mu}( \dot\Psi^{\nu} + \Psi^{\beta}\eta^{\alpha}~ \widetilde{\hat\Gamma}~^{\nu}_{.  \beta \alpha}),
\eqno(74a)$$ with $\dot\Psi^{\nu}\edf
\Psi^{\nu}_{.\hat{,}\beta}\zeta^{\beta}$. Varying (74a) with respect
to $\Psi^\sigma$ we get the following variational derivatives
$$\frac{\partial {L_{FAP}}_{\stackrel{(B)}{-}}}{\partial \psi^\sigma}=\hat{g}_{\mu\nu}\zeta^\mu \zeta^\alpha \widetilde{\hat\Gamma}~^\nu_{.\sigma\alpha},$$
$$\frac{d}{d\tau^-}(\frac{\partial {L_{FAP}}_{\stackrel{(B)}{-}}}{\partial \dot{\psi}^\sigma})=\hat{g}_{\mu\sigma\hat{,}\nu}\zeta^\mu \zeta^\nu+\hat{g}_{\mu\sigma}\frac{d \zeta^\mu}{d\tau^-}.$$

Using the Euler-Lagrange equation,
 $$ \frac{d}{d\tau^{-}}(\frac{\partial{L_{FAP}}_{\stackrel{(B)}{-}} }{\partial  \dot{\Psi}^{\sigma}})- \frac{\partial{L_{FAP}}_{\stackrel{(B)}{-}} }{\partial \Psi^{\sigma}}= 0,
$$we get after some rearrangements

  \begin{equation}
 \frac{d{\zeta}^{\alpha}}{d\tau^-}+ {\hat{\cs{\mu}{\nu}{\alpha}}}
 \zeta^{\mu}\zeta^{\nu} =0.
 \end{equation}
This is the path equation corresponds to the dual connection
${\widetilde{\hat\Gamma}}~^\nu_{.\sigma\alpha}$ and characterized by
the parameter $\tau^-$ and the tangent $\zeta^\alpha$.

  \item[(iv)] \underline{The use of ${\hat{\cs{\mu}{\nu}{\alpha}}}$}\\
Theorem 3 shows clearly that it is always possible to get a Finsler space associated with the FAP. Then, one can define Cartan  linear connection (13) using the metric (41).
Consequently, one can define the Lagrangian.
    \begin{equation}\label{}
{L_{FAP}}_{(B)}= \h{i}_{\mu}\h{i}_{\nu} W^{\mu}\frac{D
\Psi^{\nu}}{Ds}
\end{equation}
 where $W^\alpha(\edf \frac{dx^\alpha}{ds}) $ is the unit tangent to the path.
 This Lagrangian is identical to (61) with $W^\alpha = U^\alpha$. So, the variation of (77) will give the path equation.

  \begin{equation}
 \frac{d{W}^{\alpha}}{ds}+ {\hat{\cs{\mu}{\nu}{\alpha}}}
W^{\mu}W^{\nu} =0.
 \end{equation}
\end{description}

 An important notice on the Bazanski approach  is that the linear connections are written explicitly in the general Lagrangian (66) using explicit definitions for $\frac{D \psi^{\nu}}{D\tau}$ . This represents
 one of the differences between the conventional and the Bazanski approaches. This point will be more discussed in the following Section.

 We can summarize the results obtained in Subsection (4.2) in the following
 table:\\
 \newpage
\begin{center}
{\bf Table 2: Coefficients of different terms in the path equations}
\end{center}
\begin{center}
\begin{tabular}{|c|c|c|c|} \hline
 ~~~~~~~~Connection used~~~~~~~~&${\hat{\cs{\mu}{\nu}{\alpha}}}$ coefficient& $\Lambda_{\mu
 \nu}^{..~\alpha}$ coefficient &Equation No. \\
 \hline
& & &\\
  ${{\hat\Gamma}^{\mu}_{.~\alpha \nu}} $ & 1&  1  &  (70)  \\
& && \\
\hline
& & &\\
   ${{\hat\Gamma}^{\mu}_{.~ (\alpha \nu )}}$ &1 &  $\frac{1}{2}$ & (73) \\
   & & &\\

\hline
& & &\\
  ${\hat{\tilde{\Gamma}}^{\mu}_{.~\alpha\nu }},~{\hat{\cs{\mu}{\nu}{\alpha}}}$ & 1&  0    &  (76), (78) \\
& && \\

\hline
\end{tabular}
\end{center}

%=============================================  OLD SECTION   ====================

\section{Discussions and Concluding Remarks}
Path equations are of great important in the geometrization philosophy. They are usually used to describe physical trajectories of test particles in geometric field theories. For example, it is shown that, in the context of GR, geodesic and null geodesic paths, of Riemannain geometry, represent successfully the physical trajectories of test particles and of photons,  respectively.

In this piece of work, we have derived the path equations for the type of geometry, suggested in 2009,
 known as the FAP geometry. The equations are derived using the Bazanski approach and conventional approach for comparison.
 Also, path equations in Finsler geometry are derived using the Bazanski approach. The treatment
 in the present work shows that:
 \begin{enumerate}
   \item The use of the Bazanski approach in Finsler geometry (Section 3) gives rise to the same path equations obtained using the conventional
   approach. The main advantage of using the Bazanski approach is that one can obtain path deviation equation using the same Lagrangian
   (61). This is the subject of a future work in preparation now.
   \item In deriving the path equations in the FAP-geometry, we have used the Cartan type connection (27), although other connections may be
   defined in this geometry. This is because (27) is a metric connection, i.e. paths characterized
    by this connection  preserve  the metric. This property is an important one in physical applications.
   \item In Section 4 we derived path equations in the FAP-geometry using the conventional approach and the Bazanski approach. Comparison between
   the two approaches shows that:
   \begin{description}
     \item[(i)] For the horizontal path equations the results are different. The conventional approach gives only
     one equation (65) while the Bazanski approach gives three different equations (70), (73) and (76).
      One of these equations is that obtained from the conventional approach (equation (78)). This means that the Bazanski
      approach covers the conventional one and gives more results.
     \item[(ii)] A striking feature appeared when using the Bazanski approach to derive the horizontal path equations.
      This is illustrated in the third column of Table 2. The coefficient of torsion term jumps, from equation to the next, by
      a step of one half. This is tempting to believe that the FAP-geometry admits some quantum features. Although we have used four
       linear connections, we got only three jumping coefficients. This may be of physical interest.
   \end{description}
   \item In addition to other advantages, the Bazanski approach has the advantage of probing the existence of
   such quantum properties in geometries. It can be easily deduced that neither Riemannian  nor Finsler (Section 3)
   admits such properties. It has been shown that the AP-geometry and Einstein  non-symmetric geometry admit quantum features[5].
    In the present work we show that the FAP, also, implies such properties. The following Table gives a comparison between the h-path equations in some geometries using the two different
    approaches.\\
\end{enumerate}

%\begin{landscape}
\begin{center}
{\bf {\bf Table 3: Horizontal path equations in different
geometries}}
\end{center}

 {\small\begin{tabular}{|c|c|c|c|c|c|}
  \hline
  Geometry& Connection &\multicolumn{2}{|c|}{ Conventional   Approach}  & \multicolumn{2}{|c|}{Bazanski Approach}\\
   && Number of Eqs & Quantum Features & Number of Eqs & Quantum Features \\\hline
Riemannian [11] &S&1&No&1&No\\\hline Finsler [10]
&S&1&No&1&No\\\hline AP [1,3], [9] &NS&1&No&3&Yes\\\hline FAP [2],
[7] &NS&1&No&3&Yes\\\hline Einstein NS[13], [5]
&NS&1&No&3&Yes\\\hline
\end{tabular} }
\begin{description}
  \item[~~~~]
  In this Table S means symmetric and NS means non-symmetric.
From this table it is clear that path equations with quantum features are associated with non-symmetric
linear connections of the geometry considered. Such features cannot be explored using the conventional approach but using the Bazanski
one. In other words it is associated with the existence of non-vanishing torsion in the geometry.
\end{description}

%\end{landscape}

\begin{description}
  \item[5.] Let us discuss a point appeared at different positions in Section 4.2. Why the Bazanski approach gives more path equations than the conventional one, in some cases? The reason is that the form of the Lagrangian function, e.g. (66), used in the Bazanski approach contains a direct effect of the linear connection(s) admitted by the geometry. In other words the Lagrangian used contains covariant derivative of the deviation vector which is defined using the linear connection. This connection may be symmetric or non-symmetric with non-vanishing torsion. Consequently, the effect of torsion appears in the resulting path equations. It seems ( see Table 3, [10] ) that the torsion may be the cause of quantization of space time paths. This may through some light on a different philosophy for quantizing gravity.

  Fig. 1 gives a schematic comparison between path equations in different geometries.
\end{description}

%\begin{landscape}
\begin{displaymath}
    \xymatrix{ \fcolorbox{black}{white}{\begin{tabular}{c} {\bf FAP} \\    Three horizontal path\\ equations  \\ \end{tabular}
   } \ar[dd]^{\hat C_{[\alpha\beta]\gamma}=0}_{Theorem ~3} \ar[r]^{\C{i}_{\mu \nu}=0}_{Theorem~ 2}&   \fcolorbox{black}{white}{\begin{tabular}{c} {\bf AP} \\ Three path equations \\~ \\\end{tabular}
  }\ar[dd]_{g_{\mu\nu}\edf\hh{i}_\mu \hh{i}_\nu}\\&\\
    \fcolorbox{black}{white}{\begin{tabular}{c} {\bf Finsler} \\ One horizontal path \\equations  \\ \end{tabular}}\ar[r]^{ \bf C_{\alpha \beta \gamma}=0}_{Theorem ~1}& \fcolorbox{black}{white}{\begin{tabular}{c} {\bf Riemannian} \\ One path equation \\(Geodesic)\\\end{tabular}}}
\end{displaymath}
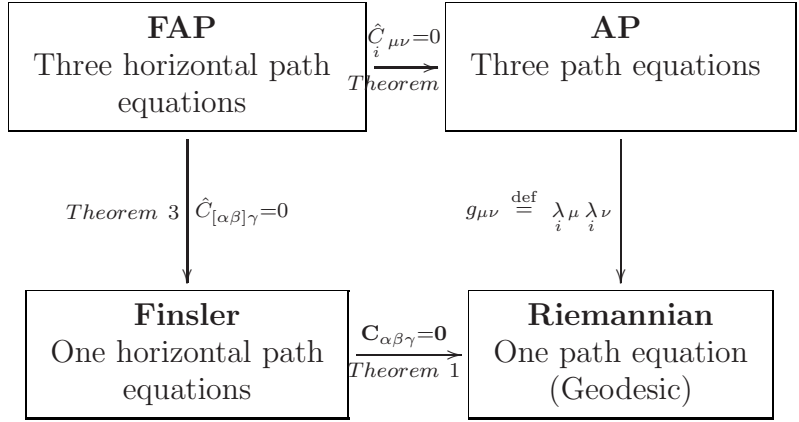
\begin{center} Figure 1: Relations between path equations in different geometries using the Bazanski approach \end{center}
%\end{landscape}

\begin{description}
  \item[6.] The present article, together with the previous two articles [2], [7], represent the main
  skeleton of the FAP-geometry. Now this type of geometry is ready for application.  The anti-curvature tensor (55) is an important geometric object, which cannot be defined in any other geometry except in the AP-geometry and its different versions.
  Its importance appears in cosmological applications of field theories  especially when talking the problem of accelerating expansion of the Universe [14]. Similarly, the W-tensor of the type given by (56-60) cannot be defined in other geometry, but the Ap-geometry  and its different versions. This tensor is important in constructing field theories [15], [16]. Many Ap-structures have been constructed for different applications cf. [17], [18]. The construction of field theories in the FAP domain and FAP structures for applications need many efforts, in progress now.
\end{description}

%\section*{References}

}
\end{document}